# Hot Electron Generation and Manipulation in Nano-spiked Plasmonic Cavity Arrays

Siddhartha Banerjee[1, *] and Jolly Xavier[1, *]

[1] SeNSE, Indian Institute of Technology Delhi, Hauz Khas, New Delhi -110016

* Corresponding author: idz228535@iitd.ac.in, jxavier@sense.iitd.ac.in

## ABSTRACT

The generation of the localized surface plasmon resonance (LSPR) on the surface of plasmonic structures in the nanoscale has paved the way for advanced biosensing, surpassing the conventional detection limits. The electric field enhancement (electromagnetic hot spots) between two plasmonic nano structures at close quarters produces hot electrons with a change in the electron density of the material. Techniques such as photo-injection are suitable to inject hot electrons from the metal crossing the Schottky barrier to the semiconductor leading to the development of photocurrent. "Sea-Urchin" looking spiked nanoparticles serve a great interest in research of plasmonic materials as they can generate hot electrons at higher rate compared to other structures. By means of finite element method (FEM) analysis, we investigate the electric field enhancement generated between the cavity of nano-star spike pairs and extend the study to an array of them. We further do a parametric study and identify the importance of using such an array to generate hot electrons whose applications are envisaged for light harvesting, enhanced photodetection, photo catalysis etc.

## 1 Introduction

Plasmonic hot carriers have been an active part of the research field in the solid-state physics and have varied application such as in the generation of "clean energy" by using a solar cell[1]–[3] , photophysical method for water splitting using visible light[4], [5], and also in biological sensing[6]. Hot electrons'(Hot-Es) uses can be seen in the field of nuclear sciences where in a controlled environment for a shock ignition of nuclear fuel for nuclear fusion process, which is better option than a traditional laser heating [7].

It has been theorized classically that the generation of Hot-Es can be seen in solid, liquid and gases and in its extensive study considered electrons being scattered by atoms and their momentum and energy transfer in the same process. This is a highly structure dependent action in case of a liquid but not in case of a solid[8]. The creation of Hot-Es is primarily seen in places near the enhancement of local electric field or "hotspots" by the means of light confinement inside a plasmonic nanocrystal or between two evenly spaced plasmonic material[9][10], [11].

In the quantum interpretation of such a phenomenon of Hot-E, the generation of the same can be potentially executed when the non-conserved linear momentum of highly excited electrons in a confined small space comes into the theoretical modelling picture. This swaying away from the conservation laws is directly proportional to the nonuniform electric field that are generated at these enhanced spots[12].

One might get confused with surface plasmon resonance (SPR) and Hot-E generation process as both are light activated phenomena. Plasmonic resonance is induced when light of certain frequency falls on the plasmonic materials such as nano stars and gets the surface electrons excited such that they oscillate between the negative and the positive permeability material with a propagation constant β [13].

$$\beta = k * \sqrt{\frac{\epsilon_o * \epsilon_m}{\epsilon_o + \epsilon_m}}$$

Where k is the incident wave propagation vector, $\epsilon_o$ is the permittivity of the dielectric and $\epsilon_m = \epsilon_{m-Real} + \epsilon_{m-Imaginary}$ are the permittivity of the metal. This is generated entirely on a surface for a given period and at a particular angle[14]. When these materials are in proximity, the modes that travelled before on the surface of these materials, now get confined at a spot and lead to localized surface plasmon resonance or LSPR. But Hot-Es have a little different characteristic than these optically activated electrons generating SPR or LSPR. Hot-Es tend to absorb the energy of the photon and reach a higher energy level than electrons reached by SPR effect and tend to stay for a small moment. The rate of production of Hot-Es as per the quantum approach is by considering one particle density matrix for a single nanoparticle can be defined with the following equations[12]

$$\delta \bar{n}(\varepsilon) = \frac{1}{\sqrt{\pi}\delta_R} \int_{-\infty}^{\infty} dR \cdot e^{-\left(\frac{R-R_o}{\delta_r}\right)^2} \cdot \delta n(\varepsilon, R)$$

Where $\tau_e$ is the characteristic relaxation time for energy ε, $E_f$ is the Fermi level, δε represents the broadening of the energy of the individual term, $\delta \bar{n}$ is the energy distribution function, $\delta_R$ is the parameter describing the size variation of the plasmonic material, $R_o$ is the average size of each of that particular material(for example a plasmonic nanosphere), and δn(ε) is defined as population of the energetic states given by

$$\delta n(\varepsilon) = 2 \sum_n \delta \rho_{nn} \cdot P(\varepsilon - \varepsilon_n)$$

Where factor 2 is responsible for showing the spin of the electron, $\rho_{n,n}$ being the wave function representing many-body particles of an electronic system to the one particle density matrix and $P(\varepsilon - \varepsilon_n)$ is the pulse function which collapses when $|\varepsilon| > \frac{\delta\varepsilon}{2}$ and becomes equal to $\frac{1}{\delta\varepsilon}$ when $|\varepsilon| < \frac{\delta\varepsilon}{2}$. In case of a dimer formed by two nanoparticles, the HoT-E are generated due to the presence two energy states that are separated by $\hbar\omega$. Also, the quantum effects are constrained to size of a nanoparticle and decreases as the size increases[12].

Periodicity of such a structure has been previously investigated and been used in applications such as lasing[10][15], non-linear optics[16], [17] etc. Hence in our study we have simulated and designed periodic arrangement of plasmonic nanoparticles (i.e. nano stars/ nano "urchins" (NU)) in each lattice to look into the effect of the change in parameters such as change in the distance of the tip of two NU etc. The structure of this paper is arranged as follows: First we investigate the numerical techniques used in detail using the FEM based COMSOL Multiphysics and next discussion of the results and parameters involved and finally concluding our results.

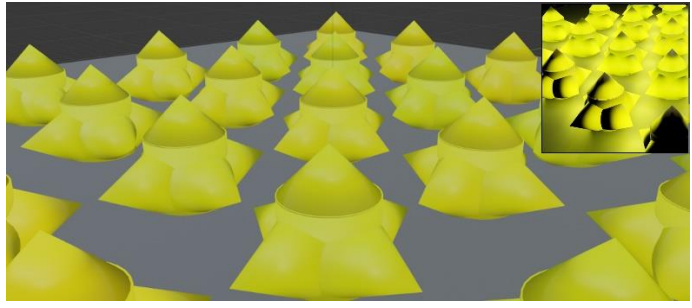

Figure 1: Schematic of the periodic array of spiked nano-particles (made using Blender 3D open source) and the inset image shows the illumination of the nanoparticle under close proximity depicting LSPR.

## 2 Computational Modelling

Our study is based on a dual approach. Firstly, we look into the three-dimensional model and then the two-dimensional structure of the same. The schematic of the model is given in figure 1 that describes nano-urchins (NU) in a periodic lattice. We have used a three-dimensional model in which we have considered the Electromagnetic Waves, Frequency Domain. The parameters were appropriately chosen as shown in the above table and the structure formed in both the x and y axis. The model was made by first considering a large sphere like a gold nano seed and the cones oriented in the x and y axis with a small sphere attached to the tip to give a smoothening effect to the entire structure. The periodic condition was placed on all the boundaries of this block and the top and bottom used as the input and output periodic port with the light impinging normally in the TE mode. A similar approach was taken in the case of the 2D simulation with the same study being considered.

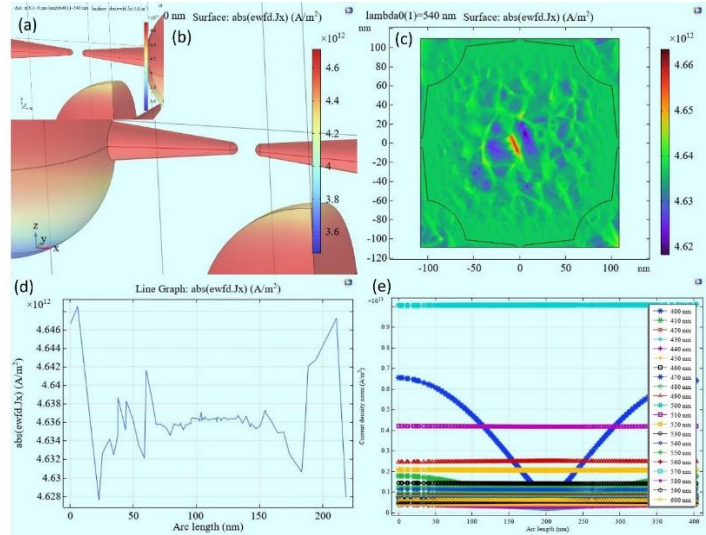

Figure 2: (a-b) Three-dimensional model of the proposed NUs showing the absolute surface current density on an incident field of 540nm of TE mode light(a) Showing the close-up view of the variation of absolute surface charge density (c) Absolute Surface charge density of the X-Y plane using the 2D cut slice at a position of z=0 nm. (d) The absolute surface current density component in the x axis varying with the arc length joining the center of the sphere of the x axis to the other large sphere in the same axis as shown in (b).(e) Variation of Normal surface charge density vs the arc length connecting the center of the Nano seed

# 3 Results and Discussion

In the three-dimensional model, the chosen incident wavelength in transverse electric (TE) mode is 540 nm. This was an initial assumption as the absorption of light is highest in the range of 500 to 700 nm for the particle size varying in the range of 20 to 100 nm[19]. Figure 2a shows the zoomed tips of the NUs tips and figure 2b shows a variation in the overall geometry. Figure 2c is the variation of the same in XY plane axis at z=0 nm. Figure 2d is the variation of the absolute surface charge density with the variation of arc length considered from the center of one large sphere to the another. This was done to record the variation of the surface charge density fluctuation from the tip to that of the center of the plasmonic nano-seed. As we record them, we see spikes of the fluctuation of the surface charge and the amplitude of the spikes grows higher from the tip of NUs to the nano-seed. This variation in the surface charge can be the generation of non-uniform electric field[12] and the LSPR generating at the tip of these materials.

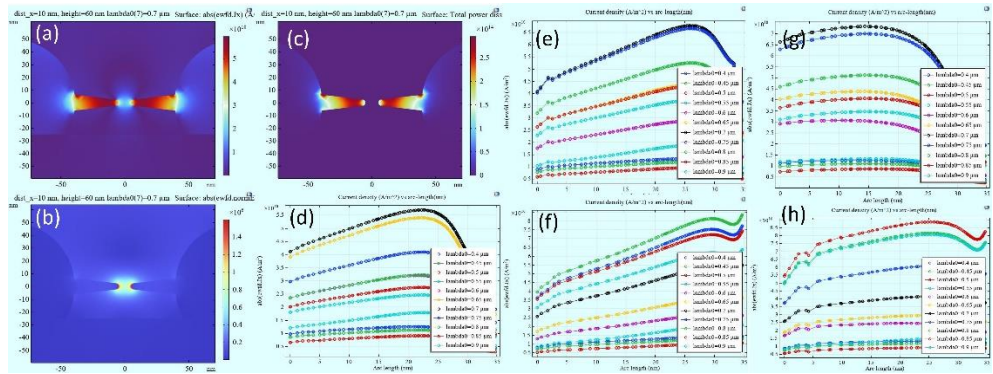

Figure 3: (a-c)This is the two dimensional analysis of the gold tip at 0.7 micrometer of TE excitation incident normally from the top. (a)The surface charge density of the x component. (b) The normalized E field at the center of the gold tips of the NUs. (c) The total power dissipation from the structure. (d-f) This is the variation of the absolute current density for a cone height of 60 nm taken from the tip of the cone to the center of the nano-seed. (g)-(h) This is the variation of the absolute current density for a cone height of 120 nm taken for the same length as above.

Next, we shift our focus to the variation of wavelength for a fixed cone length of 60 nm and 10 nm in x axis and 10 nm of y axis from the tips of the NUs. The variation is shown in figure 2 (e). The surface charge density peaks at 500 nm with a value approximately $10^{13}$ A/m$^2$. Hence local field enhancement at this incident wavelength is the highest.

Next, we find the results of the two-dimensional structure of the same and investigate the parametric change in the height of the cone and also the distance between the tips in the x axis. Figure 3d to 3f shows the variation of incident wavelength and the amount of charge density increasing for different distance of tips from each. For 0.7 micrometer of incoming wavelength with TE polarization, the highest surface charge density is seen for a cone height of 60 nm and tip distance of 10 nm and with the tips distance decreasing that is the tips move closer to each other, the shift to higher wavelength is observed. In figure 3g and figure 3h, similar results can be seen for a cone height of 120 nm and distance between tips as 10 nm and 2 nm respectively.

## 4  Conclusions

In the results discussed above one can see the variation of the surface current density on the structure of the gold nano-urchins (NUs). From the results, it is observable that the two-dimensional simulation is off by a factor of 100 in calculation of surface charge density and hence can be considered as a coarse result compared to the three-dimensional structure study. The red shift in the excitation wavelength on the surface charge density suggests that with increasing proximity between the NUs, the energetic electrons density on the surface increases. This can be a direct inference that the excitation is providing a series of high kinetic energy Hot-Es to be released which can be further studied with a time response simulation that can be a future scope of this project.

## 5  Acknowledgements

SB gratefully acknowledges "DST-INSPIRE" (DST/INSPIRE/03/2022/001266) Fellowship from Department of Science and Technology, India.